# Quantification of the effect of mutations using a global probability model of natural sequence variation

Thomas A. Hopf[1,2,*], John B. Ingraham[1,*], Frank J. Poelwijk[3], Michael Springer[1], Chris Sander[4], Debora S. Marks[1,§]

[1] Department of Systems Biology, Harvard Medical School, Boston, Massachusetts, USA
[2] Bioinformatics and Computational Biology, Department of Informatics, Technische Universität München, Garching, Germany
[3] Green Center for Systems Biology, UT Southwestern Medical Center, Dallas, Texas, USA
[4] Computational Biology Center, Memorial Sloan Kettering Cancer Center, New York, New York, USA

[*] Joint first authors
[§] Corresponding author (debbie@hms.harvard.edu)

## Abstract

Modern biomedicine is challenged to predict the effects of genetic variation. Systematic functional assays of point mutants of proteins have provided valuable empirical information, but vast regions of sequence space remain unexplored. Fortunately, the mutation-selection process of natural evolution has recorded rich information in the diversity of natural protein sequences. Here, building on probabilistic models for correlated amino-acid substitutions that have been successfully applied to determine the three-dimensional structures of proteins, we present a statistical approach for quantifying the contribution of residues and their interactions to protein function, using a statistical energy, the evolutionary Hamiltonian. We find that these probability models predict the experimental effects of mutations with reasonable accuracy for a number of proteins, especially where the selective pressure is similar to the evolutionary pressure on the protein, such as antibiotics.

## Introduction

A major challenge facing modern biology and medicine is to predict the effects of genetic variation. Genetics regularly identifies specific mutations that give rise to phenotypic effects and recent technology using large-scale mutational scans has allowed the systematic exploration of the phenotypic landscapes one mutation away from natural protein sequences[1-14]. Despite the large scale of these approaches, laboratory experiments are restricted to a small fraction of the potential mutational space and their interpretation depends critically on the particular functional properties assayed. However, natural evolution plausibly has performed a large number of mutational experiments to identify those sequences that retain sufficient function[15], and the record of these natural experiments is becoming increasingly accessible by modern genomic sequencing efforts. This rich information resource of sequence variation under evolutionary constraints provides an unprecedented opportunity to develop quantitative and predictive computational methods for linking genotype to phenotype in molecular detail, one of the hitherto elusive goals of evolutionary biology. Here, we present a general computational approach for predicting the effects of mutations on protein function in terms of interactions between residues in the protein. The approach is based solely on the sequences present in families of evolutionarily related proteins and builds in part on the successes of similar statistical models for determining the three dimensional structures of proteins[16-20].

Recent work that applies statistical models of epistatic constraints, called evolutionary couplings, in protein families, to accurately predict three-dimensional contacts in protein structures[16-23] suggests that epistatic effects, via residue-residue interactions, can be systematically identified from natural sequence variation. These global models of epistasis in proteins – in contrast to models that treat each sequence position independently - have been applied to predict the functionality of virus variants and the thermostability of small protein domains[24,73] which suggests that they might also be applied more generally to capture the phenotypic effects of mutations on proteins[25].

Previous computational approaches do not systematically take into account basic and transitive inter-residue dependencies[26-31], even if they consider epistatic interactions to some extent [30,32]. The customary independence assumption is made despite evidence suggesting that the effects of mutations depend on context throughout protein evolution [11,34-39]. These context dependencies were elegantly demonstrated in recent work showing that human disease alleles, which are present in other organisms, can be rescued by just one large compensatory mutation relative to the human gene[32]. Lastly, many previous computational methods are trained on disease association data subject to errors and ascertainment bias or rely largely on evolutionary conservation for their predictive capacity.[13,33]

The model presented here is based solely on natural sequence variation between iso-functional members of protein families, for which data is being obtained at a rapidly increasing rate. To test the ability of our model derived from evolutionary sequence data to serve as an overall predictor of the fitness effects of protein sequence variation and to provide insight into the role of particular interactions, we compare the statistical energies for mutant sequences with the experimental performance in laboratory assays from systematic mutation scans in 13 protein families, using available datasets[1-14,40]. Our analysis includes reasonably accurate prediction of experimentally determined higher order mutations many higher order mutations and the of the effects of thousands of variants of the human influenza nucleoprotein[41] on thermostability, and of hemagglutinin on viral fitness[40], which also indicates the power of using across-species variation for understanding human influenza protein evolution

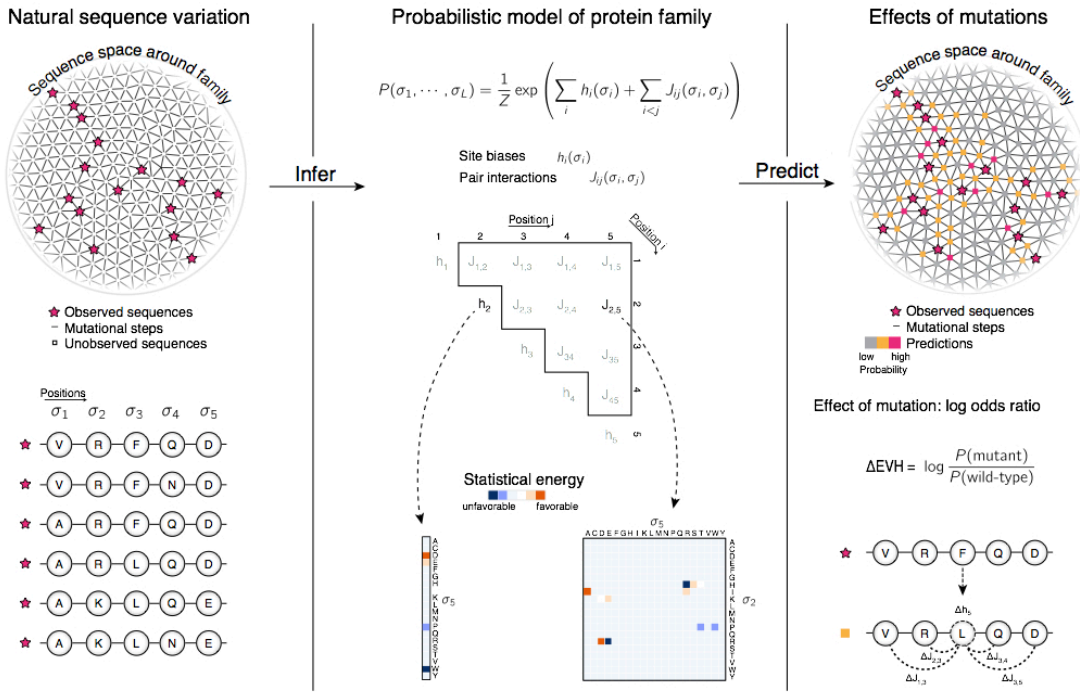

**Figure 1. Predicting context-dependent effects of mutations by modeling the distributions of protein families over sequence space. (a)** A family of related protein sequences is a set of points within a small region of sequence space. By fitting a statistical model to the constrained patterns of amino acid usage present in observed sequences, a global statistical model can predict a distribution of functional sequences between and around the observed. **(b)** *Left:* Sequences that have been under selection to maintain a particular function over evolution will show constrained patterns of amino acids usage *Middle:* The distribution of the protein family over sequence space is parameterized by a combination of site-specific bias parameters $\mathbf{h}_i$ and pairwise epistatic constraints $\mathbf{J}_{ij}$. Each $\mathbf{h}_i$ is a vector unique to each position (column) in the family that describes the relative favorability of different amino acids at that positions, while each $\mathbf{J}_{ij}$ is a matrix unique to each pair of positions describing an interaction pattern for the relative favorability for different combination of amino acids at those positions. The values of these parameters are selected to maximize the probability of observing the natural sequences, with additional penalties for model complexity. *Right:* The effects of mutations can be regarded as predictions that a given mutation will stay in the functional space of the family and can thus be quantified by log-odds scores under the probability model. For a point mutation to one protein, the log-odds score will be the difference in a the affected site parameters $\mathbf{h}_i$ for the mutated site plus the difference in all couplings $\mathbf{J}_{ij}$ that extend from the mutated site outwards.

## Probability model for sequences in a protein family

The fundamental quantities in the model are the probability for a given protein sequence to arise as a member of a particular protein family, the corresponding statistical energy (logarithm of the probability) and, importantly, the full set of specific residue interactions that contribute to the energy (Fig.1). Each protein family is modeled as a distribution over all of sequence space that is parameterized by two types of constraints: site specific terms that reflect an average bias for a particular position to favor particular amino acids, and pairwise interaction terms that bias pairs of positions to favor particular combinations of amino acids. The form of the distribution can be interpreted under the principle of maximum entropy as the least structured distribution that is consistent with the observed protein family sequence data up to second order, i.e., with the counts (marginal distributions) of amino acids at single positions and pairs of positions in a sequence alignment (Fig. 1). The probability of a protein sequence $\sigma$ under the exponential model is defined as

$$P(\sigma) = \frac{1}{Z} \exp(E(\sigma))$$

The statistical energy $E(s)$ for a particular sequence $\sigma$ is the sum of all its residue pair couplings $J_{ij}$ and single residue terms (fields) $h_i$, where $i$ and $j$ are residue positions along the sequence:

$$E(\sigma) = \sum_{i=1}^{N} \mathbf{h}_i(\sigma_i) + \sum_{i=1}^{N-1} \sum_{j=i+1}^{N} \mathbf{J}_{ij}(\sigma_i, \sigma_j)$$

For convenience, we refer to the statistical energy as an evolutionary Hamiltonian (EVH) and the probability model as the EVH model. We estimate the parameters in the probability model for sequences in the family with an approximate penalized maximum likelihood framework based on pseudo-likelihood [22,23,42-44] (Methods). Using the inferred probability distribution over sequence space for a given protein family, one can quantify the effect of a single mutation or higher-order mutation on a particular sequence background by computing the log-odds ratio of probabilities, or statistical energy difference, between the normal ('wild-type') and the mutant sequences (Fig. 1):

$$\Delta E(\sigma^{(\mathrm{mut})}, \sigma^{(\mathrm{wt})}) = E(\sigma^{(\mathrm{mut})}) - E(\sigma^{(\mathrm{wt})}) = \log \frac{P(\sigma^{(\mathrm{mut})})}{P(\sigma^{(\mathrm{wt})})}$$

For comparability between different models, the log-odds scores were normalized on a per-protein basis. Hence the effect of a mutation incorporates the sequence context into which mutations are introduced by an explicit sum over all interaction energies (Fig.1) and thereby incorporates epistatic effects as well as cooperative effects that emerge from the set of pair interactions in an equilibrium statistical ensemble.

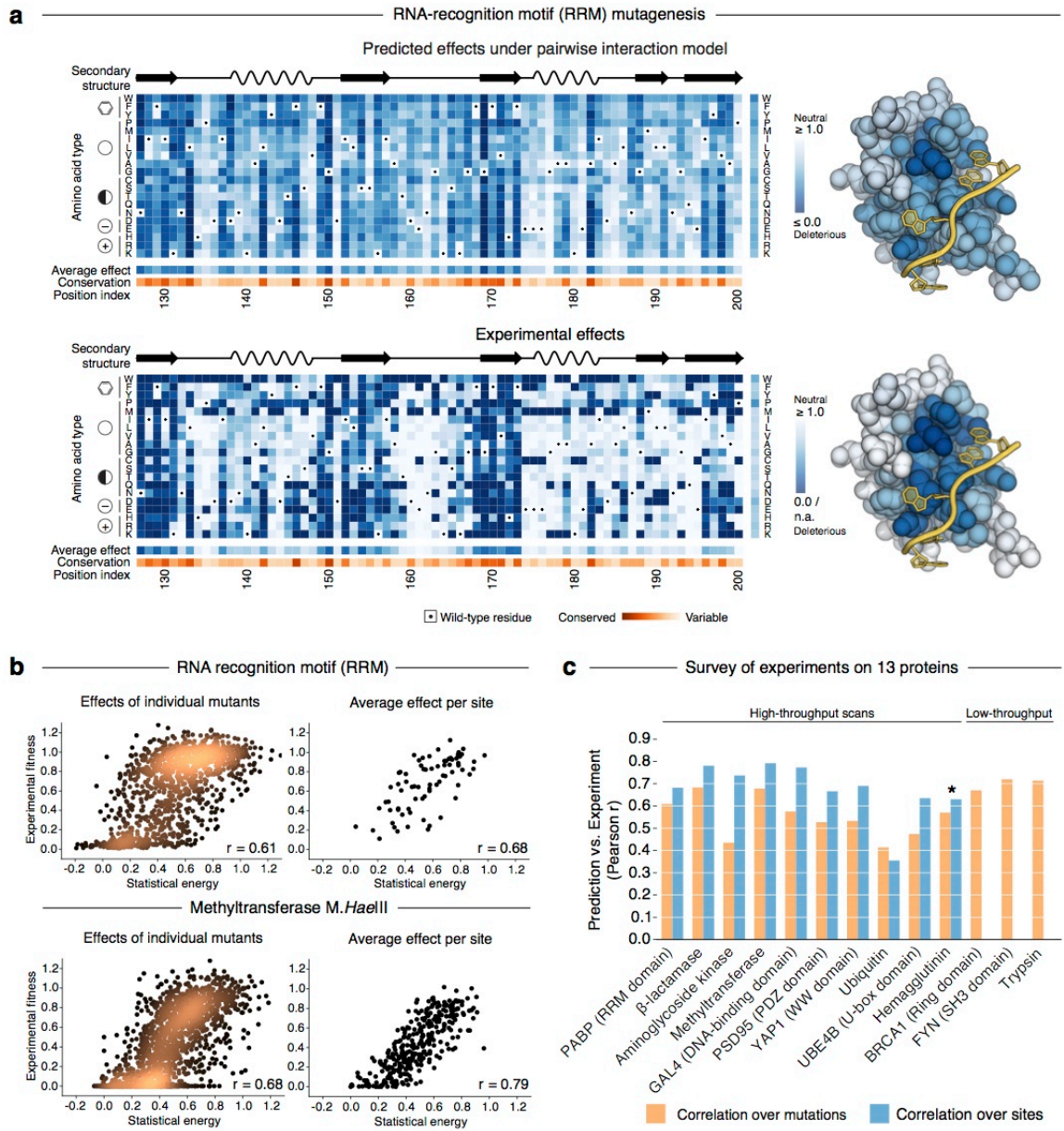

**Figure 2. Predicted effects of mutations effects correlate with measured phenotypic effects. (a)** *Left:* EVH-predicted (top) and experimentally measured *in vivo* (bottom) effects of individual point mutations to an RNA recognition motif (RRM) domain of the yeast poly(A)-binding protein[3] (x-axis: RRM sequence, y-axis: amino acid substitutions, color: effect of mutation (white: neutral, bue: deleterious). *Right*: Positional average effect of all 19 possible substitutions mapped on the structure of human PABP (RNA ligand in yellow). **(b)** Statistical energies correlate with experimentally measured growth phenotypes of RRM (top panel) and the methyltransferase M.*Hae*III (bottom panel) on the level of individual mutations (left) and when averaging over all substitutions per site (right). Some outliers outside displayed ranges not explicitly shown here..(**c**) Correlation of EVH predictions with experimentally measured mutant effects for a set of 13 tested proteins. (*) Performance for hemagglutinin was assessed on evolutionarily observed substitutions only (3340 of 11280 mutations).

## Predictions of effects of mutations

Given the inferred protein family-specific EVH models from evolutionary sequence information, we assessed the extent to which models derived from evolutionarily related sequences across many species can be used to predict the assay performance or fitness for synthetic sequences. We collected data from published saturation mutagenesis experiments[1-9,11,13,14,40] (Extended Data Table 1) and computed all possible pair and single residue terms for each protein (Supplementary Data Set 1) and then compared the experimental effects for synthetic variants to the computed changes in the statistical energy (EVH).

The predicted mutation effects correlate well, but not perfectly, with the experimentally measured outcomes of the functional assays (Fig. 2c, Extended Data Figs. 1 and 2, Extended Data Table 2). For instance, the effect predictions of 34,745 single and double mutations of the yeast RNA-binding protein PABP positively correlate with experimental effects[3] (r=0.61, average per site, *r*=0.68; Fig. 2a,b; Extended Data Fig. 3a) and fitness effects of 1685 mutations of the bacterial DNA methyltransferase M.*Hae*III correlate positively with the effects seen in selection experiments[14] (r=0.68 for all mutations, r=0.79 for average effect for each site). The high throughput experiments mostly measure sequence enrichments as a proxy for fitness, and since many mutations that alter the protein function alter thermostability [45,46], we also analyzed sets of low-throughput *in vitro* measurements on melting temperatures[47,48] (Fig. 2d). The change in EVH, statistical energy, of sequences for 47 mutations in the SH3 domain of human protein Fyn and 22 mutations in rat trypsin-2 capture the variation in melting temperatures ($T_m$) with high accuracy (r=0.72 and 0.71 respectively) (Fig. 2d). These results are consistent with evolutionary de-selection of instability and suggesting that the model could be used to predict stability changes from sequence information, at least for some residue positions

As the EVH statistical model provides a statistical energy for any sequence, one can predict the effect of any number of mutations. It is therefore possible to assess parts of the protein family landscape that have not been explored by natural variation and ask how likely natural evolution can access these through a viable chain of single mutations. We explored the simplest example of an evolutionary path in two proteins and asked how many of the predicted viable (Methods) sequences with double mutations are accessible by one or more viable paths. For all the proteins, one–path only double mutations are dominated by a relatively small number of specific amino acid changes in step 1 that are predicted to allow a unfavorable mutation in step 2. In beta lactamase the most enabling step 1 mutations are the well-known so-called global suppressor M180T and a known stabilizing mutation, N50A [45,49,50]. Similarly, in the PABP protein, we identify G177E and L202Q as global suppressors in step 1 allow over 50 otherwise deleterious mutations, many of which are in RNA-binding residues and are distant from the enabling mutation (G177E previously reported as an enabling mutation [51] (Extended Data Fig. 3b and Data Supplement (web), see Methods for URL). It will be interesting to explore the potential of the EVH method to map out evolutionary mutational pathways. Overall, the results in this section indicate that EVH predictions of assay outcomes in mutational scans can be used for experiment design, for the exploration of evolutionary dynamics or as a rapid first estimate of expected mutational effects before costly experiments.

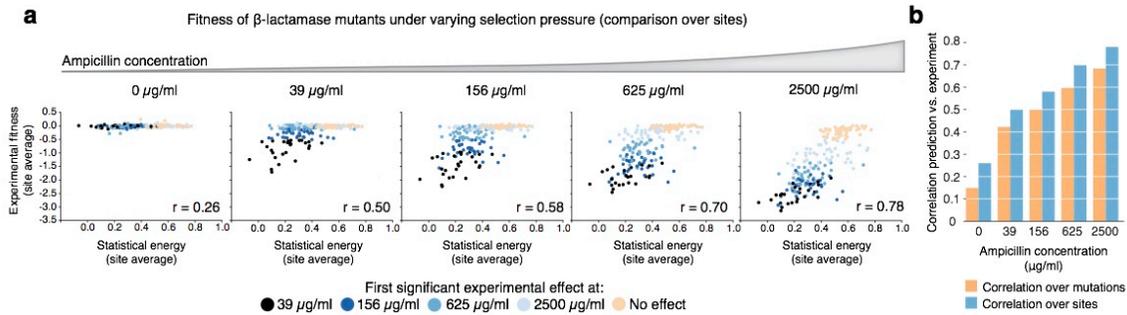

**Figure 3. The correlation between predicted and measured effects sharpens with increasing selection pressure. (a)** Many mutations to TEM-1 β-lactamase which are predicted deleterious by the model are revealed as deleterious *in vivo* by increasing selective pressure through higher ampicillin concentrations (average effect per position; left to right, shades of blue additionally indicate concentration of first significant effect determined by fitting a two-component Gaussian mixture model). **(b)** The amount of phenotypic variation explained by the predictions increases under increasing strength of antibiotic selection.

### Especially good model predictions for particular selection experiments

Experiments that have assessed the mutational effects with different types and levels of selection pressure on the same protein provide an opportunity to compare the signals from evolutionary sequence variation, which reflect the contributions of a protein to natural selection, to those from the laboratory assays, which focus on a particular functional property of a protein that is not necessarily crucial in evolution.

Sets of experiments that have assessed the mutational effects with different types and levels of selection pressure on the same protein provide an opportunity to compare the change in the EVH to across the different experiments. For instance, a series of saturation mutagenesis experiments on β-lactamase[4] and a bacterial kinase[8] (that targets aminoglycosides), measured the effects of a range of antibiotic pressures on bacterial growth. In the former, as the dose of the antibiotic ampicillin is increased, the predicted effects of mutations correlate increasingly well to the experimental fitness effects of more than 4000 mutations (from r=0.15 to r=0.68 for specific effects; r=0.26 to r=0.78 for average effects per site) (Fig. 3a, b). In the latter, the lowest kanamycin dose gives the widest range of experimental effect and the best correlation with computed effects (r=0.74 for average per site) and as selection pressure is increased and the effect saturates, the correlation with the computed EVH changes decreases (Extended Data Fig. 3c). When we combine the experimental data on the bacterial kinase from 8 different antibiotics the correlations improve – possibly suggesting that the selective pressure on the enzyme has been under a group of these antibiotics.

These examples and others such as thermostability of SH3 and trypsin versus binding or kinetics, show the tendency of the predictions, to correlate best with assay when the selection pressure in the experiment is most relevant to the *in vivo* functional contribution of the proteins to processes essential for positive selection or avoidance of de-selection.

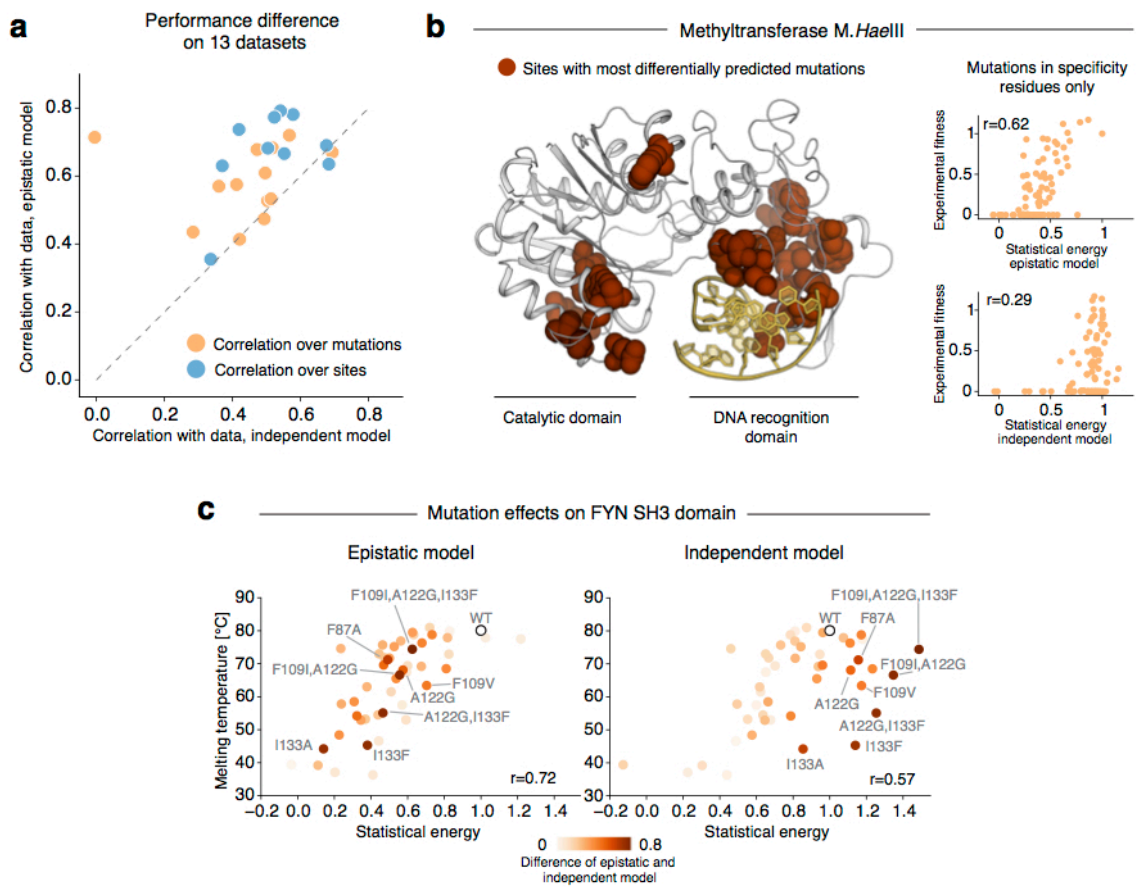

**Figure 4. Incorporating epistatic interactions systematically improves predicted effects of mutations. (a)** For many of the analyzed 13 datasets, predicted effects from the epistatic co-evolution model agree more closely with experimental measurements than those from the independent model. **(b)** The effects of single, double and triple mutants to the melting temperature of Fyn SH3 are captured more accurately by the epistatic model (r=0.72), while several destabilizing mutations are predicted as neutral by the independent model (r=0.57). **(c)** Mutations to the methyltransferase M.*Hae*III that are most differentially predicted between the epistatic and independent model cluster around the DNA recognition and catalytic domains (open conformation, PDB ID: 3ubt). When predicting the effects of mutations for specificity-determining residues, incorporating epistatic interactions leads to considerably better agreement with the experimental data (r=0.62, top right inset) than predicting effects independent of sequence context (r=0.29, bottom right inset).

### Interactions in epistatic models essential for good predictions

For the experiments on the 13 proteins analyzed here, the epistatic EVH model predicts experimental effects more accurately than an equivalent 'independent' statistical model that does not consider inter-dependencies between sites, i.e., has no interaction parameters $J_{ij}$ in the Hamiltonian (Fig. 4a, Extended Data Figure 5). This suggests that the epistatic, EVH model, which has built in interaction terms, is a better descriptor of family-specific constraints on protein sequences. To explore in which way the epistatic EVH probability model improves predictions, we considered the subset of experiments that that had the largest difference in predictive performance with and without pair interaction terms: M.*Hae*III, β-lactamase, GAL4, bacterial kinase and PABP protein (Fig. 4b, Extended Data Fig. 5). The experimentally unfavorable mutants are predicted more differentially than all other mutations (Extended Data Fig. 6a). For most of the proteins the improvement of the full EVH model over the independent model is substantial for mutations of residues involved in binding or active site (Fig. 4b, r=0.61 for the epistatic model vs. *r*=0.28 for the independent model), and improvement of the epistatic model is particularly pronounced for the set of residues that define *interaction* specificity (Extended Data Table 3).

The effects of single and higher-order mutations on melting temperature are well captured by the epistatic model for SH3 and trypsin (r=0.72 and r=0.71 respectively) but very poorly for the independent model., (Extended Data Table 2). For SH3 and trypsin, incorporating pairwise epistatic constraints accounted for a 20% and 45% increase (respectively) to the explained variation in melting temperatures In SH3, these mutants involve co-dependent substitutions to four residues (F87, F109, A122 and I133) that form a contiguous network of predicted interactions within the core. Here, the independent model fails to predict the deleterious effect on melting temperatures (Fig. 4c). Similarly, mutations of residues M109 and C160 to alanine in trypsin result in large reductions in the melting temperature (~15°C) that are only accurately predicted by the epistatic model. Although alanine is frequently observed in other sequences in the protein family alignment, i.e. in a different background, it is deleterious in the background of the target sequence. Taken together this suggests that considering context-dependence between sites responsible for functional specificity is crucial for capturing sequence constraints and deleterious mutation effects in these positions.

## Discussion

Inferring epistatic models with all pairwise residue-residue interactions presents a difficult statistical challenge, since the typical number of free parameters (~$10^6$-$10^8$) vastly exceeds the number of available sequences (~$10^3$-$10^5$). This discrepancy requires the identifying a reasonable tradeoff between the complexity of the model and the fit to the observed data (Methods), which may be improved with both the acquisition of more sequence data as well as more sophisticated approaches for model regularization. With the potential for overfitting in mind, expectation of the accuracy of computational predictions should naturally consider experimental noise, as the correlation of the biological replicates in some of the experiments presented here is ~ r = 0.7 which puts an upper bound of the extent to which we can expect to explain variation in assayed properties.

Our probabilistic approach, based only on sequence variation, can be used to study specific variants and combinations thereof for numerous protein families. We suggest that future analyses of genetic variation and inquiries into the mechanisms of molecular evolution will benefit from global probability models of sequence families that explicitly incorporate interactions between positions such as presented here. The increased accessibility of genomic sequencing will enable this unique opportunity to turn the evolutionary sequence record into a quantitative and predictive resource for connecting genotype to phenotype, not only for proteins.

## Methods

**Mutation effect datasets**

Mutation effect datasets were identified by a comprehensive literature search for quantitative high-throughput mutagenesis experiments of entire proteins or protein domains. All experiments that targeted proteins with insufficient sequence diversity (redundancy-reduced number of sequences <10$L$, where $L$=length of protein or domain) were excluded from the final compilation of datasets (Supplementary Table 1). For further comparisons, the dataset was extended with low-throughput measurements of protein function for SH3 and trypsin, with a focus on sequence co-evolution studies.

**Multiple sequence alignments**

For each protein in our dataset compilation (query sequence), multiple sequence alignments of the protein family were obtained by five search iterations of the profile HMM homology search tool jackhmmer[58] against the UniRef100 database of non-redundant protein sequences[59]. To control for comparable evolutionary depth across different protein families, we used length-normalized bit score sequence similarity thresholds as described previously[23]. A default bit score of 0.5 bits/residue was used as the threshold for minimum sequence similarity unless the alignment did not result in ≥80% coverage of the length of the query domain or if there were not enough sequences; in the former case, the threshold was increased in steps of 0.05 until sufficient coverage was obtained; in the latter case, the threshold was decreased until there were sufficient sequences (redundancy-reduced number of sequences ≥10$L$). The resulting alignments were post-processed to exclude positions with more than 30% gaps and to exclude sequences that align to less than 50% of the length of the query sequence. To adjust for the variable density sampling of sequence space by evolution, sequences were reweighted by the inverse of their number of similar neighbors at an 80% identity cutoff as described previously[18,19].

**Inference of epistatic statistical model of sequences**

To capture in full generality the variable site- and pair-specific constraints on amino acid sequences in specific protein families, we applied the maximum entropy principle to model each family as a distribution over sequence space constrained by the pairwise empirical marginal distributions of amino acids at each pair of positions. The form of this distribution can be thought of as the least-structured (i.e. highest-entropy) global distribution over sequence space that is consistent with the single-site and pairwise marginal distributions of amino acids observed in the alignment[19,61-63]. Under this model, the probability of any arbitrary amino acid sequence $\sigma$ of length $N$ is defined as

$$P(\sigma) = \frac{1}{Z} \exp\left( \sum_{i=1}^{N} \mathbf{h}_i(\sigma_i) + \sum_{i=1}^{N-1} \sum_{j=i+1}^{N} \mathbf{J}_{ij}(\sigma_i, \sigma_j) \right) \quad (0.1)$$

where the partition function $Z$ sums over all possible sequences $\sigma'$ to ensure correct normalization of the distribution. The site specific parameters $\mathbf{h}_i(\sigma_i)$ and pair specific parameters $\mathbf{J}_{ij}(\sigma_i,\sigma_j)$ collectively implement the constraints that the marginal probability distributions of the model agree with the empirical marginal frequencies $f_i(\sigma_i)$ and $f_{ij}(A_i, A_j)$. This class of exponential probability models is commonly known as Markov random field[64], or 20-state Potts model in statistical physics[65,66].

Treating a reweighted multiple sequence alignment as a set of approximately independent samples from a protein family, the model parameters $\mathbf{h}_i$ and $\mathbf{J}_{ij}$ could in principle be estimated by maximizing their likelihood, i.e. by finding the set of parameters that maximizes the probability of the observed sequence data. This is however intractable due to the $20^N$ summations over all of sequence space in the partition function $Z$. As a replacement for the full likelihood function, a site-factored pseudolikelihood maximization (PLM[43,44]) approximation is used[42], which is a consistent estimator in the limit of large data. PLM has been applied previously to protein families for residue-residue contact prediction with high accuracy and is used as default estimator on the evfold.org web server.

The number of parameters for the model, which for typical protein families of 50-500 amino acids will range from $10^5$ to $10^7$, vastly outnumbers the typical number of sequences available for even the largest families, which ranges from $10^2$ to $10^5$. In this under-sampled regime, standard maximum likelihood estimation is highly prone to overfit the sample data. To ensure that the model generalizes to unobserved sequences, a penalty for model complexity was added in the form of $l_2$-regularization[22,44]. This form of penalized maximum likelihood estimation may also be interpreted as maximum *a posteriori* inference under zero mean-Gaussian priors on the model parameters with a variance equal to half the inverse of lambda. Parameters for $l_2$-regularization were set as $\lambda_h$=0.01 for the field parameters $\mathbf{h}_i$ and $\lambda_J$=0.2(N-1) for the coupling parameters $\mathbf{J}_{ij}$.

**Calculation of evolutionary couplings**

Between any two positions *i* and *j* in the protein family, the coupling matrix $J_{ij}$ describes the relative favorability of all possible $20^2$ amino acid combinations. To summarize these sequence-specific parameters into a single description of the total epistatic constraint between pairs, a summary statistic was computed as the root sum of squares of the matrix elements (Frobenius norm). After computing the norm scores for every pair of positions, background modes of constraint caused by limited sampling and phylogenetic relationships between sequences were removed by subtracting site-averaged correction terms (average product correction)[67]. The top-ranking set of significantly constrained pairs, referred to as evolutionary couplings, was selected by a previously described cutoff determined from the noise distribution[23].

**Context-dependent prediction of mutation effects (Evolutionary Hamiltonian)**

Due to the Boltzmann form of the distribution, the relative likelihood of any particular sequence in a protein family can be quantified by its statistical energy, i.e. the sum of field and coupling parameters that contribute to its relative weight in the probability distribution. For any particular sequence *σ*, the statistical energy (from here on called Evolutionary Hamiltonian, EVH) is defined as

$$E(\sigma) = \sum_{i=1}^{N} \mathbf{h}_i(\sigma_i) + \sum_{i=1}^{N-1} \sum_{j=i+1}^{N} \mathbf{J}_{ij}(\sigma_i, \sigma_j) \tag{0.2}$$

In the chosen sign convention, higher statistical energies are more favorable and correspond directly to higher probabilities.

Mutation effects were quantified by the difference in statistical energy between the mutant sequence $\sigma^{(mut)}$ and the wild-type sequence $\sigma^{(wt)}$, which may also be interpreted as the log-odds ratio of probabilities of the two sequences under the inferred model:

$$\Delta E(\sigma^{(\mathrm{mut})}, \sigma^{(\mathrm{wt})}) = E(\sigma^{(\mathrm{mut})}) - E(\sigma^{(\mathrm{wt})}) = \log \frac{P(\sigma^{(\mathrm{mut})})}{P(\sigma^{(\mathrm{wt})})} \qquad (0.3)$$

For a given mutant, the difference in statistical energy will be the sum of differences of the field parameters for all mutated sites, plus the sum of differences of the coupling parameters for all pairs of positions involving at least one mutated site. Values of $\Delta E$ above 0 correspond to more probable mutant sequences (beneficial mutation), values below 0 to less probable mutant sequences (deleterious mutation) and values equal to 0 to equally probable sequences (neutral mutation).

Since statistical energy differences are not directly comparable between different proteins, a rescaling based on the most deleterious predicted effects ($D$, estimated by the mean statistical energy difference of the 5% most deleterious single mutants to reduce the influence of single outliers) is applied to calculate the final mutation effect as

$$\Delta E_c(\sigma^{(\mathrm{mut})}, \sigma^{(wt)}) = \frac{\Delta E(\sigma^{(\mathrm{mut})}, \sigma^{(wt)})}{|D|} + 1 \qquad (0.4)$$

After rescaling, $\Delta E_c = 1$ corresponds to neutral mutations and $\Delta E_c = 0$ to strongly deleterious mutations. Predicted effects of mutations based on Equation (1.1) are denoted as from the "epistatic model".

**Inference of independent maximum entropy model**

To assess the information gained by using an epistatic probability model of sequence evolution, additional simpler maximum entropy models were inferred that describe protein sequences only with site-specific amino acid preferences $\mathbf{h}_i$, without considering inter-dependencies between sites (i.e., predict mutation effects *independent* of the sequence context). The probability of any amino acid sequence $\sigma$ under this model is given by

$$P(\sigma) = \frac{1}{Z} \exp\left(\sum_{i=1}^{N} \mathbf{h}_i(\sigma_i)\right) \qquad (0.5)$$

Consistent with the regularization applied to the pairwise epistatic model (Equation 1.1), the strength of the $l_2$ penalty was set at $\lambda_h$=0.01 when estimating the model parameters **h**$_i$. The statistical energy difference between two sequences can be analogously inferred by substituting the single-site model probabilities into Equation (1.3) and rescaling with Equation (1.4). Mutation effects computed based on the single-site model are denoted as from the "independent model". This formalism is related to the conservation-based features used in many methods to predict mutation effects from sequence[29][30].

**Quantification of the context-dependence of mutations**

The proportion of a predicted mutation effect that can be attributed to context-dependence between sites was quantified by calculating the difference between the predicted effects $\Delta E_c$ (rescaled log-odds ratios) from the epistatic and independent models:

$$\Delta\Delta E_{\text{ind}}^{\text{epi}}(\sigma^{(\text{mut})}, \sigma^{(\text{wt})}) = \Delta E_c^{\text{epi}}(\sigma^{(\text{mut})}, \sigma^{(\text{wt})}) - \Delta E_c^{ind}(\sigma^{(\text{mut})}, \sigma^{(\text{wt})}) \qquad (0.6)$$

The more negative or positive the value of $\Delta\Delta E_{\text{ind}}^{\text{epi}}$ is, the more the effect of a mutant sequence variant depends on the sequence background, according to the two models.

**Classification of experimental mutation effects**

Based on the observation that the effect distribution of many experimental mutation scans is bimodal, mutation effects were classified as deleterious or neutral/low effect by fitting two-component Gaussian mixture models to the data. Mutation effects (enrichment ratios of sequencing reads before and after functional selection) were transformed into log-space where applicable. Individual mutants were then classified by assignment to the mixture model component returning the higher posterior probability.

**Analysis of structural features**

Evolutionary couplings calculated from multiple sequence alignments were compared to experimental protein 3D structures from the PDB[68] to assess if the identified epistatic constraints correspond to structural contacts. Two residues are considered to be in contact if any of their atoms are closer than 5 Å; a distance threshold of 4 Å was applied to interactions between amino acid residues and ligands. Mappings between UniProt sequences and PDB structures were obtained from the SIFTS database[69] and jackhmmer-based alignments.

**Data analysis and availability**

All data analysis was conducted using IPython notebooks[70] and the scientific Python stack[71,72]. The Extended Data Figures and Tables, Supplementary Data and an example IPython notebook are available at the website marks.hms.harvard.edu/evh/. All software and code will be made publicly available upon publication of the paper.

Note added in proof: The Weigt lab has performed a related analysis with comparable conclusions on the role of epistatic couplings in predicting mutational effects[74]

# References

1  Roscoe, B. P. & Bolon, D. N. Systematic exploration of ubiquitin sequence, E1 activation efficiency, and experimental fitness in yeast. *Journal of molecular biology* **426**, 2854-2870, doi:10.1016/j.jmb.2014.05.019 (2014).
2  Roscoe, B. P., Thayer, K. M., Zeldovich, K. B., Fushman, D. & Bolon, D. N. Analyses of the effects of all ubiquitin point mutants on yeast growth rate. *Journal of molecular biology* **425**, 1363-1377, doi:10.1016/j.jmb.2013.01.032 (2013).
3  Melamed, D., Young, D. L., Gamble, C. E., Miller, C. R. & Fields, S. Deep mutational scanning of an RRM domain of the Saccharomyces cerevisiae poly(A)-binding protein. *Rna* **19**, 1537-1551, doi:10.1261/rna.040709.113 (2013).
4  Stiffler, M. A., Hekstra, D. R. & Ranganathan, R. Evolvability as a Function of Purifying Selection in TEM-1 beta-Lactamase. *Cell* **160**, 882-892, doi:10.1016/j.cell.2015.01.035 (2015).
5  McLaughlin, R. N., Jr., Poelwijk, F. J., Raman, A., Gosal, W. S. & Ranganathan, R. The spatial architecture of protein function and adaptation. *Nature* **491**, 138-142, doi:10.1038/nature11500 (2012).
6  Kitzman, J. O., Starita, L. M., Lo, R. S., Fields, S. & Shendure, J. Massively parallel single-amino-acid mutagenesis. *Nature methods* **12**, 203-206, 204 p following 206, doi:10.1038/nmeth.3223 (2015).
7  Findlay, G. M., Boyle, E. A., Hause, R. J., Klein, J. C. & Shendure, J. Saturation editing of genomic regions by multiplex homology-directed repair. *Nature* **513**, 120-123, doi:10.1038/nature13695 (2014).

8       Melnikov, A., Rogov, P., Wang, L., Gnirke, A. & Mikkelsen, T. S. Comprehensive mutational scanning of a kinase in vivo reveals substrate-dependent fitness landscapes. *Nucleic acids research* **42**, e112, doi:10.1093/nar/gku511 (2014).

9       Fowler, D. M. *et al.* High-resolution mapping of protein sequence-function relationships. *Nature methods* **7**, 741-746, doi:10.1038/nmeth.1492 (2010).

10      Araya, C. L. *et al.* A fundamental protein property, thermodynamic stability, revealed solely from large-scale measurements of protein function. *Proceedings of the National Academy of Sciences of the United States of America* **109**, 16858-16863, doi:10.1073/pnas.1209751109 (2012).

11      Podgornaia, A. I. & Laub, M. T. Protein evolution. Pervasive degeneracy and epistasis in a protein-protein interface. *Science* **347**, 673-677, doi:10.1126/science.1257360 (2015).

12      Firnberg, E., Labonte, J. W., Gray, J. J. & Ostermeier, M. A comprehensive, high-resolution map of a gene's fitness landscape. *Molecular biology and evolution* **31**, 1581-1592, doi:10.1093/molbev/msu081 (2014).

13      Starita, L. M. *et al.* Massively Parallel Functional Analysis of BRCA1 RING Domain Variants. *Genetics*, doi:10.1534/genetics.115.175802 (2015).

14      Rockah-Shmuel, L., Toth-Petroczy, A. & Tawfik, D. S. Systematic Mapping of Protein Mutational Space by Prolonged Drift Reveals the Deleterious Effects of Seemingly Neutral Mutations. *PLoS Comput Biol* **11**, e1004421, doi:10.1371/journal.pcbi.1004421 (2015).

15      Harms, M. J. & Thornton, J. W. Evolutionary biochemistry: revealing the historical and physical causes of protein properties. *Nature reviews. Genetics* **14**, 559-571, doi:10.1038/nrg3540 (2013).

16      Marks, D. S., Hopf, T. A. & Sander, C. Protein structure prediction from sequence variation. *Nature biotechnology* **30**, 1072-1080, doi:10.1038/nbt.2419 (2012).

17      Hopf, T. A. *et al.* Three-dimensional structures of membrane proteins from genomic sequencing. *Cell* **149**, 1607-1621, doi:10.1016/j.cell.2012.04.012 (2012).

18      Morcos, F. *et al.* Direct-coupling analysis of residue coevolution captures native contacts across many protein families. *Proceedings of the National Academy of Sciences of the United States of America* **108**, E1293-1301, doi:10.1073/pnas.1111471108 (2011).

19      Marks, D. S. *et al.* Protein 3D structure computed from evolutionary sequence variation. *PLoS One* **6**, e28766, doi:10.1371/journal.pone.0028766 (2011).

20      Jones, D. T., Buchan, D. W., Cozzetto, D. & Pontil, M. PSICOV: precise structural contact prediction using sparse inverse covariance estimation on large multiple sequence alignments. *Bioinformatics* **28**, 184-190, doi:10.1093/bioinformatics/btr638 (2012).

21      Ovchinnikov, S., Kamisetty, H. & Baker, D. Robust and accurate prediction of residue-residue interactions across protein interfaces using evolutionary information. *eLife* **3**, e02030, doi:10.7554/eLife.02030 (2014).

22      Kamisetty, H., Ovchinnikov, S. & Baker, D. Assessing the utility of coevolution-based residue-residue contact predictions in a sequence- and structure-rich era. *Proceedings of the National Academy of Sciences of the United States of America* **110**, 15674-15679, doi:10.1073/pnas.1314045110 (2013).

23   Hopf, T. A. *et al.* Sequence co-evolution gives 3D contacts and structures of protein complexes. *eLife* **3**, doi:10.7554/eLife.03430 (2014).

24   Lapedes, A., Giraud, B. & Jarzynski, C. Using sequence alignments to predict protein structure and stability with high accuracy. *arXiv preprint arXiv:1207.2484* (2012).

25   Shekhar, K. *et al.* Spin models inferred from patient-derived viral sequence data faithfully describe HIV fitness landscapes. *Physical review. E, Statistical, nonlinear, and soft matter physics* **88**, 062705 (2013).

26   Reva, B., Antipin, Y. & Sander, C. Predicting the functional impact of protein mutations: application to cancer genomics. *Nucleic acids research* **39**, e118, doi:10.1093/nar/gkr407 (2011).

27   Sim, N. L. *et al.* SIFT web server: predicting effects of amino acid substitutions on proteins. *Nucleic acids research* **40**, W452-457, doi:10.1093/nar/gks539 (2012).

28   Bromberg, Y., Yachdav, G. & Rost, B. SNAP predicts effect of mutations on protein function. *Bioinformatics* **24**, 2397-2398, doi:10.1093/bioinformatics/btn435 (2008).

29   Adzhubei, I., Jordan, D. M. & Sunyaev, S. R. Predicting functional effect of human missense mutations using PolyPhen-2. *Current protocols in human genetics / editorial board, Jonathan L. Haines ... [et al.]* **Chapter 7**, Unit7 20, doi:10.1002/0471142905.hg0720s76 (2013).

30   Reva, B., Antipin, Y. & Sander, C. Determinants of protein function revealed by combinatorial entropy optimization. *Genome Biol* **8**, R232, doi:10.1186/gb-2007-8-11-r232 (2007).

31   Bloom, J. D. & Glassman, M. J. Inferring stabilizing mutations from protein phylogenies: application to influenza hemagglutinin. *PLoS Comput Biol* **5**, e1000349, doi:10.1371/journal.pcbi.1000349 (2009).

32   Jordan, D. M. *et al.* Identification of cis-suppression of human disease mutations by comparative genomics. *Nature*, doi:10.1038/nature14497 (2015).

33   Grimm, D. G. *et al.* The evaluation of tools used to predict the impact of missense variants is hindered by two types of circularity. *Human mutation* **36**, 513-523, doi:10.1002/humu.22768 (2015).

34   Weinreich, D. M., Lan, Y., Wylie, C. S. & Heckendorn, R. B. Should evolutionary geneticists worry about higher-order epistasis? *Current opinion in genetics & development* **23**, 700-707, doi:10.1016/j.gde.2013.10.007 (2013).

35   Pollock, D. D., Thiltgen, G. & Goldstein, R. A. Amino acid coevolution induces an evolutionary Stokes shift. *Proceedings of the National Academy of Sciences of the United States of America* **109**, E1352-1359, doi:10.1073/pnas.1120084109 (2012).

36   Pollock, D. D. & Goldstein, R. A. Strong evidence for protein epistasis, weak evidence against it. *Proceedings of the National Academy of Sciences of the United States of America* **111**, E1450, doi:10.1073/pnas.1401112111 (2014).

37   Breen, M. S., Kemena, C., Vlasov, P. K., Notredame, C. & Kondrashov, F. A. Epistasis as the primary factor in molecular evolution. *Nature* **490**, 535-538, doi:10.1038/nature11510 (2012).

38  Kryazhimskiy, S., Dushoff, J., Bazykin, G. A. & Plotkin, J. B. Prevalence of epistasis in the evolution of influenza A surface proteins. *PLoS genetics* **7**, e1001301, doi:10.1371/journal.pgen.1001301 (2011).
39  Olson, C. A., Wu, N. C. & Sun, R. A comprehensive biophysical description of pairwise epistasis throughout an entire protein domain. *Current biology : CB* **24**, 2643-2651, doi:10.1016/j.cub.2014.09.072 (2014).
40  Thyagarajan, B. & Bloom, J. D. The inherent mutational tolerance and antigenic evolvability of influenza hemagglutinin. *eLife* **3**, doi:10.7554/eLife.03300 (2014).
41  Gong, L. I., Suchard, M. A. & Bloom, J. D. Stability-mediated epistasis constrains the evolution of an influenza protein. *eLife* **2**, e00631, doi:10.7554/eLife.00631 (2013).
42  Besag, J. Statistical analysis of non-lattice data. *The statistician*, 179-195 (1975).
43  Balakrishnan, S., Kamisetty, H., Carbonell, J. G., Lee, S. I. & Langmead, C. J. Learning generative models for protein fold families. *Proteins* **79**, 1061-1078, doi:10.1002/prot.22934 (2011).
44  Ekeberg, M., Lovkvist, C., Lan, Y., Weigt, M. & Aurell, E. Improved contact prediction in proteins: using pseudolikelihoods to infer Potts models. *Physical review. E, Statistical, nonlinear, and soft matter physics* **87**, 012707 (2013).
45  Tokuriki, N. & Tawfik, D. S. Stability effects of mutations and protein evolvability. *Curr Opin Struct Biol* **19**, 596-604, doi:10.1016/j.sbi.2009.08.003 (2009).
46  Bloom, J. D., Labthavikul, S. T., Otey, C. R. & Arnold, F. H. Protein stability promotes evolvability. *Proceedings of the National Academy of Sciences of the United States of America* **103**, 5869-5874, doi:10.1073/pnas.0510098103 (2006).
47  Di Nardo, A. A., Larson, S. M. & Davidson, A. R. The relationship between conservation, thermodynamic stability, and function in the SH3 domain hydrophobic core. *Journal of molecular biology* **333**, 641-655 (2003).
48  Halabi, N., Rivoire, O., Leibler, S. & Ranganathan, R. Protein sectors: evolutionary units of three-dimensional structure. *Cell* **138**, 774-786, doi:10.1016/j.cell.2009.07.038 (2009).
49  Wang, X., Minasov, G. & Shoichet, B. K. The structural bases of antibiotic resistance in the clinically derived mutant beta-lactamases TEM-30, TEM-32, and TEM-34. *J Biol Chem* **277**, 32149-32156, doi:10.1074/jbc.M204212200 (2002).
50  Huang, W. & Palzkill, T. A natural polymorphism in beta-lactamase is a global suppressor. *Proceedings of the National Academy of Sciences of the United States of America* **94**, 8801-8806 (1997).
51  Melamed, D., Young, D. L., Miller, C. R. & Fields, S. Combining natural sequence variation with high throughput mutational data to reveal protein interaction sites. *PLoS genetics* **11**, e1004918, doi:10.1371/journal.pgen.1004918 (2015).
52  Towler, W. I. *et al.* Analysis of BRCA1 variants in double-strand break repair by homologous recombination and single-strand annealing. *Human mutation* **34**, 439-445, doi:10.1002/humu.22251 (2013).
53  Suzuki, Y. & Nei, M. Origin and evolution of influenza virus hemagglutinin genes. *Molecular biology and evolution* **19**, 501-509 (2002).

54	Neher, R. A., Russell, C. A. & Shraiman, B. I. Predicting evolution from the shape of genealogical trees. *eLife* **3**, doi:10.7554/eLife.03568 (2014).
55	Luksza, M. & Lassig, M. A predictive fitness model for influenza. *Nature* **507**, 57-61, doi:10.1038/nature13087 (2014).
56	Newman, M. & Barkema, G. *Monte Carlo Methods in Statistical Physics chapter 1-4*. (Oxford University Press: New York, USA, 1999).
57	Wagner, A. Neutralism and selectionism: a network-based reconciliation. *Nature reviews. Genetics* **9**, 965-974, doi:10.1038/nrg2473 (2008).
58	Eddy, S. R. Accelerated Profile HMM Searches. *PLoS computational biology* **7**, e1002195, doi:10.1371/journal.pcbi.1002195 (2011).
59	Suzek, B. E. *et al.* UniRef clusters: a comprehensive and scalable alternative for improving sequence similarity searches. *Bioinformatics* **31**, 926-932, doi:10.1093/bioinformatics/btu739 (2015).
60	Sievers, F. *et al.* Fast, scalable generation of high-quality protein multiple sequence alignments using Clustal Omega. *Molecular systems biology* **7**, 539, doi:10.1038/msb.2011.75 (2011).
61	Jaynes, E. T. Information theory and statistical mechanics. *Physical review* **106**, 620 (1957).
62	Lapedes, A. S., Giraud, B. G., Liu, L. & Stormo, G. D. Correlated mutations in models of protein sequences: phylogenetic and structural effects. *Lecture Notes-Monograph Series*, 236-256 (1999).
63	Weigt, M., White, R. A., Szurmant, H., Hoch, J. A. & Hwa, T. Identification of direct residue contacts in protein-protein interaction by message passing. *Proceedings of the National Academy of Sciences of the United States of America* **106**, 67-72, doi:10.1073/pnas.0805923106 (2009).
64	Koller, D. & Friedman, N. *Probabilistic graphical models: principles and techniques*. (MIT press, 2009).
65	MacKay, D. J. *Information theory, inference and learning algorithms*. (Cambridge university press, 2003).
66	Landau, L. D. & Lifshitz, E. Statistical physics, part I. *Course of theoretical physics* **5**, 468 (1980).
67	Dunn, S. D., Wahl, L. M. & Gloor, G. B. Mutual information without the influence of phylogeny or entropy dramatically improves residue contact prediction. *Bioinformatics* **24**, 333-340, doi:10.1093/bioinformatics/btm604 (2008).
68	Berman, H. M. *et al.* The Protein Data Bank. *Nucleic acids research* **28**, 235-242 (2000).
69	Velankar, S. *et al.* SIFTS: Structure Integration with Function, Taxonomy and Sequences resource. *Nucleic acids research* **41**, D483-489, doi:10.1093/nar/gks1258 (2013).
70	Pérez, F. & Granger, B. E. IPython: a system for interactive scientific computing. *Computing in Science & Engineering* **9**, 21-29 (2007).
71	Van Der Walt, S., Colbert, S. C. & Varoquaux, G. The NumPy array: a structure for efficient numerical computation. *Computing in Science & Engineering* **13**, 22-30 (2011).

72  Hunter, J. D. Matplotlib: A 2D graphics environment. *Computing in science and engineering* **9**, 90-95 (2007).
73  Lui, S. & Tiana, G. The network of stabilizing contacts in proteins studied by coevolutionary data. *J Chem Phys* **139**, 155103, doi:10.1063/1.4826096 (2013).
74  Figliuzzi, M., Jacquier, H., Schug, A., Tenaillon, O. & Weigt, M. Coevolutionary landscape inference and the context-dependence of mutations in beta-lactamase TEM-1. *Mol Biol Evol*, doi:10.1093/molbev/msv211 (2015).

# Extended Data Tables

**Extended Data Table 1. Set of mutagenesis experiments.**

**Extended Data Table 2. Correlations between prediction and experiment.**

**Extended Data Table 3. Correlations between prediction and experiment for functional residues.**

# Extended Data Figure Legends

**Extended Data Figure 1. Correlations between prediction and experiment.** Scatter plots displaying the quantitative relationship between predictions from evolutionary sequence variation (epistatic and independent model) and experimentally tested phenotypes (top: individual mutants, bottom: site averages for high-throughput experiments). A single outlier was excluded from display in the scatter plots for PABP_YEAST but included in the calculation of correlation coefficients. Bacterial kinase experiments with the suffix "filtered" have data points with fitness > 5 removed. M.*Hae*III analysis was restricted to the majority of data points with single nucleotide exchanges; those with the suffix "filtered" additionally have variants with low frequencies in the initial library (≤0.01) removed.

**Extended Data Figure 2. Computational prediction of all possible single mutant effects using epistatic model.** Mutation matrix representation (see Fig. 2) of predicted single mutation effects using the context-dependent EVH model for all tested proteins. Displayed positions are those that were covered by the respective sequence alignment and included for inference of the statistical models (see Methods).

**Extended Data Figure 3. Double mutant effects in RRM domain and selective pressure on aminoglycoside kinase. (a)** Fitness effects from evolutionary sequence

information computed for a set of ~40.000 double mutants of the PABP RNA-recognition motif domain using the epistatic sequence model quantitatively correlate with experimental results more strongly (r=0.62) than those from the independent model (r=0.50). **(b)** Based on a global epistatic model of mutation effects, interactions between different mutations and mutational paths can be predicted. *Left*: Mutations G177E, G177D, L202Q enable several mutations in other sites (connected by blue arcs) that on their own would be deleterious, but are predicted neutral in the context of the enabling mutation (100 most deleterious compensated mutants shown; statistical energy ≥ 0.9; sites of enabling mutations highlighted in blue, RNA-binding residues in bold font). *Right*: Two examples of individually deleterious mutations (A179M, D136R) that are tolerated in the backgrounds of L202Q and G177E, respectively. The double mutants presumably can only be reached if the enabling mutation occurs first. **(c)** The correspondence between predicted and experimental mutation effects (site averages) for a bacterial aminoglycoside kinase depends on the applied antibiotic selection pressure. Some mutations are only revealed as deleterious *in vivo* by increasing selective pressure; overall correlations however decrease as a large fraction of mutants becomes non-viable independent of their relative fitness (left to right; shades of blue indicates concentration of first significant effect. The threshold for significant effects (positional average < 1.01) was determined by fitting a two-component Gaussian mixture model to the distribution of average site effects in log space at 1:8 WT MIC).

**Extended Data Figure 4. Correspondence of evolutionary couplings to 3D structure contacts.** Predicted epistatic pairs of positions (black dots: all pairs with evolutionary coupling score above background noise) for all tested proteins largely correspond to structural proximity of the residues in corresponding experimental protein 3D structures (5Å and 8Å minimum atom residue distance, medium and light blue dots) and define a global residue interaction network.

**Extended Data Figure 5. Difference in predicted mutation effects between epistatic and independent model.** Mutation effects for single mutants predicted using the epistatic model tend to be more deleterious than those from the independent model. For many of the tested proteins, the epistatic model effects are in better agreement with the experimental data (white: neutral mutation; blue: deleterious mutation; grey: no data available) than those by the independent model, which assigns high statistical energies to experimentally deleterious sequences since it cannot account for the context-dependence of mutations (Fig. 2c, Extended Data Table 2; dark blue data points towards top of scatter plots).

**Extended Data Figure 6. Differential predictions for high-effect mutations and specificity-determining residues.** **(a)** Mutations with high experimental effect (orange curve), as determined by fitting a two-component Gaussian mixture model, show stronger differences in predicted effects between the epistatic and independent models than all other mutations (black curve). **(b)** Mutations in specificity-determining residues (orange curve; 4Å minimum atom distance to peptide/DNA ligand or residues of interacting domains, see Methods) show stronger differences in the computed effects between the epistatic and independent models than the remaining positions (black curve).